\documentclass[twocolumn,prb,aps,superscriptaddress]{revtex4}
\usepackage{color}
\usepackage{graphicx}% Include figure files
\usepackage{subfig}
\usepackage{natbib}
\usepackage[fleqn]{amsmath}
\usepackage{amssymb}
\newcommand{\nc}{\newcommand}
\nc{\on}{\operatorname}
\nc{\wt}{\widetilde}
\nc{\Wick}{{\mathbb :}}
\nc{\R}{{\mathbb R}}

\newcommand{\beq}{\begin{equation}}
\newcommand{\eeq}{\end{equation}}
\newcommand{\bmul}{\begin{multline}}
\newcommand{\emul}{{\end{multline}}}
\newcommand\beqa{\begin{eqnarray}}
\newcommand\eeqa{\end{eqnarray}}
\newcommand\bea{\begin{array}}
\newcommand\eea{\end{array}}
\newcommand\ba{\begin{array}}
\newcommand\ea{\end{array}}

\newcommand{\neqa}{\nonumber\end{eqnarray}}

\renewcommand{\O}{{\cal O}}

\nc{\CH}{{\mathcal H}}
\nc{\Db}{{\bar D}}
\nc\comment[1]{}

\nc{\CM}{{\mathcal M}}
\nc{\CN}{{\mathcal N}}

\newcommand{\re}{\relax{\rm I\kern-.18em R}}

\nc{\meV}{{\mathrm{\,meV}}}
\nc{\cG}{{\mathcal G}}

\renewcommand{\)}{\right)}

\comment{ \)}
\renewcommand{\bar}{\overline}

\nc{\al}{{\alpha}}

\begin{document}

\title{Pechukas-Yukawa formalism for Landau-Zener transitions in the presence of external noise}
\author{Mumnuna A. Qureshi}
\address{Department of Physics and Centre for Science and Materials, Loughborough University, Loughborough LE11 3TU, UK.}
\author{Johnny Zhong}
\address{Department of Mathematical Sciences, Loughborough University, Loughborough LE11 3TU, UK.}
%\author{Zihad Qureshi}
%\address{Attero Solutions, Chatham, UK.}
\author{Peter Mason }
\address{Department of Physics, Loughborough University, Loughborough LE11 3TU, UK.}
\author{Joseph J. Betouras }
\address{Department of Physics, Loughborough University, Loughborough LE11 3TU, UK.}
\author{Alexandre M. Zagoskin}
\address{Department of Physics, Loughborough University, Loughborough LE11 3TU, UK.}

\begin{abstract}
Quantum systems are prone to decoherence due to both intrinsic interactions as well as random fluctuations from the environment. Using the Pechukas-Yukawa formalism, we investigate the influence of noise on the dynamics of an adiabatically evolving Hamiltonian which can describe a quantum computer. Under this description, the level dynamics of a parametrically perturbed quantum Hamiltonian are mapped to the dynamics of 1D classical gas. We show that our framework coincides with the results of the classical Landau-Zener transitions upon linearisation. Furthermore, we determine the effects of external noise on the level dynamics and its impact on Landau-Zener transitions. 
\end{abstract}
\maketitle

\section{Introduction}
Adiabatic quantum computers (AQC) offer an alternative to the standard approach to quantum computing, well suited for optimisation problems. One major challenge to AQC is decoherence. A generic AQC is governed by the Hamiltonian\cite{Zagoskin1, Requist4, Fahri1, Gernot, Ours,  Fahri2, Sarovar, Candia, Pechukas, Yukawa3, Haake, Requist}: %A common approach to achieve quantum coherence in non-equilibrium many body dynamics is Keldysh Green function theory\cite{Requist}, however this approximation is limited to short time intervals where its errors grow as a power of time\cite{Requist}. The study of adiabatically evolving systems are crucial to the development of adiabatic quantum computing (AQC). 

\begin{equation}
\begin{gathered}
\label{GenHam}
H(\lambda(t)) = H_0 + \lambda(t)ZH_b,
\end{gathered}
\end{equation}

\noindent where $H_0$ is an unperturbed Hamiltonian with an easily achievable nondegenerate ground state, $\lambda$ is an adiabatically evolving parameter and $ZH_b$ is a large bias perturbation term with $Z \gg 1$\cite{Berends, Huyghebaert, Zagoskin, Wilson1, Ours}. Due to the fragility of its quantum states with respect to external and internal sources of decoherence, the investigation of state transitions in adiabatically evolving systems are crucial to the development of AQC\cite{Zagoskin, Wilson1}. 

The Pechukas-Yukawa formalism maps the level dynamics of  Eq.(\ref{GenHam}) to a one-dimensional (1D) classical gas with long range repulsion\cite{Zagoskin}. It is especially convenient for AQC, taking $\lambda$ to be an adiabatically evolving parameter. An extension of the formalism describes the dynamics of the quantum states of the system\cite{Ours, Ours2} through the evolution of $C(t)$, a vector of the expansion coefficients of the quantum state over the orthonormal state of instantaneous eigenstates of the Hamiltonian. A wavefunction, expanded in the instantaneous eigenstates, is described by the following:

\begin{equation}
\begin{gathered}
\label{Wavefunction}
|\psi\rangle=\sum_n{C_n(t)|n\rangle}.
\end{gathered}
\end{equation}

The above expansion can be used to determine the density matrices, 

\begin{equation}
\begin{gathered}
\label{DensityMatrix}
\rho(t)=C(t)\bigotimes C^{*T}(t).
\end{gathered}
\end{equation}

 This provides insight on both the dynamics of occupation numbers (the probability of remaining in a state after level ``collisions'') and the coherences (inter-level correlations). 

Using the Landau-Zener (LZ) model, one assumes that the level occupation numbers only change due to LZ tunnelings at avoided level crossings (anitcrossings)\cite{Zagoskin, Kayanuma}.  The LZ probabilities detail the fundamental results of non-stationary quantum mechanics\cite{PokSin} e.g. the non-adiabatic population transfer at level crossings and anticrossings in perturbed Hamiltonian systems or quantum phase transitions\cite{Cejnar}. The LZ model has been extended  to stochastic systems\cite{PokSin}. This details the probabilities of state transitions under the influence of random environmental effects \cite{Kayanuma, Leggett}, which may lead to decoherence in the system. One source of decoherence is noise; the Landau-Zener model is suitable to describe analytically the decoherence from external noise\cite{PokSin}. % This The Landau-Zener model has been extended to stoquastic systems, detailing  These result in decoherence in a quantum system.  This describes a time dependent version of quantum tunelling with dissipation.

We develop the LZ model in the Pechukas-Yukawa formalism to gain insight on the effects of random fluctuations on the evolution of quantum states. This approach can describe a non-equillibrium interacting system of highly entangled states, especially the dynamics of a system and its vulnerability to decoherence. We investigate the compatibility of the Pechukas-Yukawa formalism and the LZ model to determine the conditions necessary for the LZ model to be applicable. We further explore the impact of noise on these requirements and the behaviour of levels approaching the point of minimum separation under the influence of noise. This paper aims at developing basic elements of such an approach, which will be especially useful for, but not necessarily restricted to, modelling adiabatic quantum computers. 

The structure of the paper is as follows: Sec. II gives a brief overview of the Pechukas equations and the evolution of the eigenstate coefficients, Sec. III provides the background of LZ transitions and its application to the Pechukas-Yukawa model. In Sec. IV, the conditions required for the applicability of the Landau-Zener model within the Pechukas-Yukawa formalism are investigated. We further outline the conditions required for the applicability of LZ approximation (isolated crossings). In Sec. V the study is extended to determine the influence of external noise on these conditions. Discussion and conclusions are presented in Sec. VII. %This was further extended to account for the impacts of dissipative influences. In Sec. IV details the conditions required for isolated separations, well suited for investigation using the Pechukas-Yukawa formalism escribing the Landau-Zener model.  %on the Landau-Zener transitions are discussed. In Sec. V we determine the effects of noise on the minimum separation of an avoided crossing. Sec. VI 

%We build on prior works by AZ and RW, which gives the potential to analytically determine the differences in the scaling factors observed between edge states and intermediate states,

%We determine the behaviour of levels approaching the point of minimum distance under the influence of noise, comparing between the effects of white noise againt those of fast and slow coloured noise. We  on the Landau-Zener transitions,  and the applicability of the Landau-Zener model under their impacts on the Landau-Zener probability distribution of remaining in the initial state after a level crossing. We compare these numerically to observe the differences between intermediate state transitions against edge state transitions. Developing a Lindbladian master equation for the evolution of quantum states under the effects of a generic noise term we determine analytically how these state transitions vary.

%extended to the effects of noise. Sec. IV we numerically determine the differences between intermediate state transitions against edges. In Sec. V we derive the evolution of the Lindbladian master equation for this system and compare occupation dynamics for intermediate states against edges.

\section{The Pechukas Model and the Evolution of eigenstate coefficients}

For completeness, we outline the approach, first developed by Pechukas, that maps the level dynamics of an externally perturbed quantum systems on that of a fictitious classical 1D gas. It is well suited though not restricted to adiabatic quantum systems\cite{Haake, Zagoskin, Ours}. The associated Hamiltonian for this gas is written as:

\begin{equation}
\begin{gathered}
\label{HamiltonianP}
H=\frac{1}{2}\sum^N_{n=1}v^2_n+\frac{1}{2}\sum^N_{n \neq m}\frac{|l_{mn}|^2}{(x_m-x_n)^2}.
\end{gathered}
\end{equation}

\noindent which is derived from the Pechukas equations, %described by the following closed set of ordinary differential equations. 
%As in Eq.(\ref{GenHam}), $H_0$ is given by the first term in the expression and $ZH_b$ by the latter. 

\begin{equation}
\begin{gathered}
\label{Pechukas}
\frac{dx_m}{d\lambda}=v_m \\
\frac{dv_m}{d\lambda}=2\sum_{m\neq n}{\frac{{{|l}_{mn}|}^2}{{(x_m-x_n)}^3}}\\
\frac{d{l}_{mn}}{d\lambda}=\sum_{k\neq m,n}{l_{mk}l_{kn}\left(\frac{1}{{(x_m-x_k)}^2}-\frac{1}{({x_k-x_n)}^2}\right)}.
\end{gathered}
\end{equation}

\noindent These equations are derived directly from quantum equations of motion for Eq.(\ref{GenHam}) using Hamilton's equations of motion, where $x_m\left(\lambda \right)=E_m(\lambda )=\left\langle m|H|m\right\rangle$, the instantaneous eigenvalues of the system, $v_m\left(\lambda \right)=\left\langle m|ZH_b|m\right\rangle $ and $l_{mn}\left(\lambda \right)=\left(E_m\left(\lambda \right)-E_n(\lambda )\right)\left\langle m|ZH_b|n\right\rangle $ which is skew-hermitian, satisifying $l_{mn}=-l^*_{nm}$. These represent the ``positions'', ``velocities'' and particle-particle repulsion as determined by the ``relative angular momenta''\cite{Pechukas, Yukawa1, Zagoskin, Ours}. Unlike the well known integrable Calogero-Sutherland model, here the "interparticle repulsion amplitudes", $l_{mn}$, are not constant and have their own dynamics. Nevertheless the system described in Eq. (\ref{Pechukas}) is also integrable\cite{Haake}. In this model, all information for the Hamiltonian dynamics is encoded in its initial condition. %Here, the derivative is with respect to $\lambda$, a time evolving parameter.  

These equations have been extended to the stochastic sense accommodating noise from random fluctuations in the environment. Using the central limit theorem; noise arises from a number of independent identical sources, therefore it is reasonable to assume that the sum of its effect is Gaussian. The contribution of the noise in the Hamiltonian is denoted through the term $\delta h(\lambda(t))$\cite{Wilson1}, $H(\lambda(t))=H_0+\lambda(t)ZH_b+\delta h(\lambda(t))$\cite{Wilson1}. For real eigenvalues, $\delta h$ is Hermitian. As simplification,  $\delta h(\lambda)$ is taken to be real. It is shown that with the added stochastic term, the Pechukas mapping still applies and we can extend Eq. (\ref{Pechukas}) to the closed stochastic Pechukas equations\cite{Wilson1}, given by the following: %, for a Brownian model describing white or a coloured spectrum. %In \onlinecite{Wilson} considered the speciffic case of the effects of coloured noise on the dynamics of the levels and its impacts on Landau-Zener tunelling, this shall be discussed further in the following chapters. describing the random nature of the stochastic system from an external noise source,

%This transforms Eq. (\ref{Pechukas}) to the following:

%\begin{widetext}
\begin{equation}
\begin{gathered}
\label{PNoise}
\dot{x}_m=v_m+\dot\delta h_{mm} \\
\dot{v}_m=2\sum_{m\neq n}{\frac{{{|l}_{mn}|}^2}{{(x_m-x_n)}^3}}+\frac{l_{mn}\dot\delta h_{nm}-\dot\delta h_{mn}l_{nm}}{(x_m-x_n)^2}\\
\dot{l}_{mn}=\sum_{k\neq m,n}{l_{mk}l_{kn}\left(\frac{1}{{(x_m-x_k)}^2}-\frac{1}{({x_k-x_n)}^2}\right)}\\
+\frac{(x_m-x_n)(l_{mk}\dot\delta h_{km}-\dot\delta h_{mk}l_{km})}{(x_m-x_k)(x_n-x_k)}+\\
\dot\delta h_{mn}(v_m-v_n)+\frac{l_{mn}(\delta h_{mm}-\delta h_{nn})}{(x_m-x_n)}.
\end{gathered}
\end{equation}
%\end{widetext}

\noindent The derivative, denoted by `.' is taken with respect to $\lambda$.  It is clear that the mapping retains its structure; whereby if ”$\delta h = 0$”  Eq.(\ref{PNoise}) reduces to Eq.(\ref{Pechukas}). The stochastic Pechukas equations, Eq. (\ref{PNoise}) is independent of any assumptions on the nature of the noise, therefore applicable to a wide range of stochastic systems\cite{Wilson1,Wilson2}. 
Using this formalism, we investigate the conditions for the applicability of the Landau-Zener model We further extend this description to explore the impacts of external noise on these conditions. % to the Pechukas-Yukawa formalism to explore the influence of noise on the non-adiabatic state transitions at a level (avoided) crossing.  %Given the dependence of Eq. (\ref{PNoise}) on the the derivative of $\delta h$, the noise term must obey a simple stochastic differential equation, Wilson considered an Ornstein-Uhlembeck process for coloured noise, however the model is not restricted in this way but takes allows for all noise models.  Turning off noise, $\delta h_{mn}= 0$, Eq.(\ref{PNoise}) reduces to Eq.(\ref{Pechukas}), retaining the key feature of an exact mapping of quantum eigenvalue dynamics to a classical gas.These equations encode the same level dynamics as the standard Pechukas equations.

%Noise provides a source of decoherence in a quantum system however, one observes decoherence in noiseless systems, a consequence of level crossings and avoided crossings resulting in state transitions\ref{Zagoskin1, Zagoskin2, Wilson1, Wilson2, Haake, Pok, Kiss, Sin}. These transitions are described by the Landau-Zener model; using the Pechukas-Yukawa formalism, we explore the effects of the noise on the model. 

%By denoting \cite{Ours}: 

%then Eq.(\ref{EvolC})  can be written as

\section{Landau-Zener Transition Probabilities}

\noindent The Pechukas equations, Eq.(\ref{Pechukas}) are well suited to describe level crossings and anticrossings in a system. Level crossings occur when $x_m(\lambda^{*})=x_n(\lambda^{*})$ describing degeneracies\cite{Vitanov, Kleppner, Wilkinson1}, as a result $l_{mn}(\lambda^{*})=0$ at some level crossing at $\lambda^*$ (converse is not necessarily true\cite{Zagoskin, Ours}). Anticrossings arise when levels approach a minimum non-zero distance before repelling. The standard approach to model the interactions assumes all other level interactions are negligible reducing the system to 2 interacting levels about $\lambda^{*}$. Anticrossings are parameterised by the size of the gap at closest approach and the asymptotic slope of the curves\cite{Wilkinson1, Wilkinson2, Zakrezewski}. For an isolated anticrossing, the energy levels take hyperbollic form: $x^{\pm}(\lambda)=x(\lambda^{*})+B(\lambda-\lambda^{*})\pm \frac{1}{2}(\Delta_{min}^2+A^2(\lambda-\lambda^*))^{\frac{1}{2}}$ with $\Delta_{min}$ denoting the minimum gap size, $B(\lambda-\lambda^{*})$  and $A(\lambda-\lambda^{*})$ respectively describing the mean and the difference in the assymptotic slopes\cite{Wilkinson1, Wilkinson2}. 

The LZ model is used to describe these interactions through a statistical distribution of gap sizes, governing the rate of excitation due to non-adiabatic population transfers. This gives the probability to remain in its initial state after a level crossing or anticrossing. For an adiabatic regime independent of external noise, this probability is given by the probability distribution\cite{Zagoskin, Amin2}:  %Using the Pechukas model, we consider a Gaussian unitary ensemble (GUE) where levels exhibit quadratic level repulsion\cite{Wilkinson1, Zakrezewski}. The probability distribution for this case is given by $P(S)dS=\alpha_\nu S^\nu dS$, where $\nu=2$, representing the order of repulsion and $\alpha=\frac{\pi^2}{3}$. This determines the theThese are explored under the Landau-Zener model. The Landau-Zener model describes

%\begin{widetext}
%\begin{eqnarray}
\begin{equation}
\begin{gathered}
\label{PLZ}
%P_{LZ}=e^{\frac{1}{4\pi\dot\lambda}\frac{\Delta_{min}^2(x_m-x_n)}{l_{mn}l_{nm}}}
%P_{LZ}=e^{\frac{-2\pi}{\lambda}\frac{|\langle m|ZH_b|n\rangle|}{\dot\lambda}}\\
P_{LZ}=e^{-\frac{\Delta^2_{min}}{4\pi|\langle m|ZH_b|n\rangle|\dot\lambda}}, 
\end{gathered}
\end{equation}
%\end{eqnarray}
%\end{widetext}

%\noindent where $\Delta_{min}$ represents the minimum separation between the levels at $\lambda^{*}$. 
\noindent The transition time, $\tau_{LZ}=\Delta_{min}/\dot\lambda$ is defined by the time interval the levels interact in a neighbourhood $\gamma$ of each other (for a level crossing this interaction is instantaneous)\cite{Vitanov, Wilkinson1, PokSin, Kieselev, Luo}. Under the Pechukas-Yukawa formalism, one can determine from the initial conditions whether a system will exhibit quantum phase transitions and their impacts on the system\cite{Cejnar}.% It was suggested in [\onlinecite{Cejnar}] that one can expect quantum phase transitions to occur if there is a fast initial compression of the whole spectrum, leading to sharp scatterings, however this investigation is beyond the scope of this paper. %Without loss of generality, phase effects are taken to be zero. We consider a system such that phase transitions do not occur. In the case of occurances such as Griffiths phase, the Landau-Zener model collapses. %We explore the likliness of multi-level crossings based on probabilistic arguments imposed on the random nature of the system.%In the non-adiabatic regime, the probability of population transfer between states at an avoided crossing between neighbouring levels is given by\cite{Amin2, Zagoskin1, Wilson1} $P_{LZ}=e^{2\pi\frac{l_{mn}l_{nm}}{\dot\lambda (x_m-x_n)^2}}$.  For the remainder of this paper, we consider an adiabatic regime with transition time; 

\section{Landau-Zener conditions on the deterministic Pechukas-Yukawa formalism}

The applicability of the LZ transition model requires that both the perturbation parameter $\lambda$ and level separation are traversed linearly in time, localised about $\lambda^*$. Furthermore, under the LZ model the $N$ level system collapses to a 2 level problem where only the interacting levels\cite{Wilkinson1, Zakrezewski} play a significant role. This comes from the assumption that level crossings are locally more dominant than all other interactions during this period such that contributions from far away levels can be neglected. %Multi-level crossings are rare events, with negligible statistical significance\cite{Zakrezewski}.
%Then taking that close to a level crossing or anticrossing, both $\lambda$ and the level separation must be a linear function of time and expanding about $\lambda^{*}$, we linearise the system. %in the neighbourhood of the interaction and the level separation evolves linearly about . %Taking a linear evolution in $\lambda$, the former condition is satisfied.  

%In order for the LZ model to work, transitions between two anticrossing levels at a given time should dominate. 
In Pechukas-Yukawa formalism, this is a plausible assumption: due to the "two-body" interactions fast decaying  with distance, the collisions are practically independent, and the influence of other "particles" is expected to be small. Our further analysis shows that this is actually the case. Furthermore, we find that the Pechukas-Yukawa formalism can indeed be simplified to linear level separations. We examine the behaviour of the level separations about $\lambda^{*}$ using a Taylor expansion. For a level crossing, we have shown the relative angular momenta terms are constantly $0$ and the acceleration terms independently tend to 0. This demonstrates linear evolution in level separations. See Appendix A for details. 
%Using the underlying assumption that levels outside the level crossing or anticrossings are far away and that their coupling interactions are by comparison negligible. We investigated the impacts on the Pechukas equations, Eqs.(\ref{Pechukas}) to show that in the Pechukas-Yukawa formalism, the $N$ level system can be reduced to solely the interacting levels.

On the other hand, anticrossings have constant relative angular momenta, $\beta$ between levels at the level crossing or anticrossing. In this case all other relative angular momenta $l_{mi}$ and $l_{ni}$, are constants where $Re(l_{mi}), Re(l_{ni})=0$ and $Im(l_{mi}), Im(l_{ni})$ are bounded in the interval $[-1, 1]$. The difference between the accelaration terms of the interacting levels is constant, $\frac{4|\beta|^2}{\Delta_{min}^3}$ at $\lambda^{*}$. Choosing $\delta\lambda$ sufficiently small, these terms are negligible therefore linearising the level separations. Details are provided in Appendix B. Under these approximations, the Pechukas-Yukawa formalism is reduced to the Calogero-Sutherland model. 

In the setting of bosonic systems, this compares with the works in [\onlinecite{Gangardt}] where the coupling constant in our system is given by the golden ratio. It was shown for coupling strengths in the interval $(1,2)$ the system can be described as a quasi-super-solid where the potential energies are of the same order as the kinetic energies. %In the case that the potential energies are much less than the kinetic energies, the system of bosons can be described by a Wigner crystal.

\subsection{Isolated Crossings}

\noindent To satisfy that the non-interacting levels can be ignored in a LZ transition, we must ensure that level crossings are isolated from each other. We compare the differences in the transition times between level crossings or anticrossings in a close vicinity of each other. Given that the transition times do not overlap, these level crossings and anticrossings can be regarded as independent of each other. %The LZ transition model is applicable when level crossings or anticrossings are independent of all other level interactions.
 
For level crossings, $\tau_{LZ} \rightarrow 0$. This reflects a strong repulsion between the levels such that the transition time is instantaneous. Given that multi-level crossings are statistically negligible and that no more than 2 levels in a close vicinity cross at a single point so the level crossings are independent of each other, we devote our attention to 2 level anticrossings occuring in a close vicinity with minimum level separations at $\lambda^{*}$ and $\lambda^{**}=\lambda^*+\delta$ and transition times $\tau_{LZ}$ and $\tau^{'}_{LZ}$ respectively as in Fig. 1. We take symmetric anticrossings such that $\tau_{LZ}=2\xi$. Recall that in the adiabatic regime, $\tau_{LZ}=\Delta_{min}/\dot\lambda$.  These anticrossings are considered isolated given that their respective transition times do not overlap such that $(\lambda^{**}-\xi^{`})-(\lambda^*+\xi) > 0$. Then, the Landau-Zener transition model is applicable to describe the probabilities of population transitions.

%\begin{widetext}
\begin{figure}
\begin{center} 
\includegraphics[width=\linewidth]{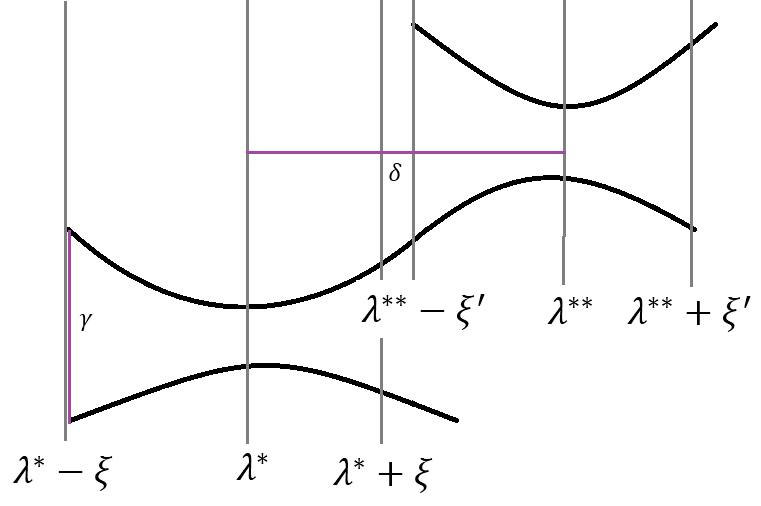}
\end{center}
 \caption{2 anticrossings in a close vicinity of each other.}
\label{fig:IsolatedCrossing}
\end{figure}
%\end{widetext}

We denote the distance between levels $d(\lambda) = x_m-x_n$, where $x_m>x_n$, $m$ and $n$ label the levels involved at an anticrossing. Levels are considered to be in an anticrossing when they are in a $\gamma$ neighbourhood of each other about a local minimum denoted by $d(\lambda^{*})=\Delta_{min}$ where $\dot d(\lambda^{*})=0$. Expanding $d(\lambda)$ about $\lambda^*$ where $\delta\lambda=(\lambda-\lambda^*)$, we obtain the following (details are provided in Appendix D):

%\begin{widetext}
%\begin{eqnarray}
\begin{equation}
\begin{gathered}
\label{ExpandGap}
d(\lambda) = \Delta_{min}+\delta\lambda^2\frac{4\beta^2}{\Delta_{min}^3}
\end{gathered}
\end{equation}
%\end{eqnarray}
%\end{widetext}

\noindent Let $d(\lambda^*+\xi)=\gamma$, then the minimum separation is expressed as $\Delta_{min}=\gamma-\xi^2\frac{4\beta^2}{\Delta_{min}^3}$. Given that  levels are within a distance $\gamma$ of each other and $\delta >\frac{1}{2\dot\lambda}(\gamma-\frac{4\beta^2}{\Delta_{min}^3})+ \xi^{'}$, the anticrossings are considered isolated and one can apply the LZ model.  %For a single source of noise, the contributions from noise cancel, reducing to the deterministic bound.
Next, we extend this investigation of the impact of noise under these conditions. This enables further understanding of dissipative influences on the properties of level interactions. % contributions from non-interacting levels are negligible, we impose that level separations between interacting levels, and non-interacting levels are negligible. Under this assumption, %Under this approximation, we determine the Pechukas equations Eqs.(\ref{Pechukas}) and (\ref{PNoise}).  

\section{The impacts of noise on the Landau-Zener conditions in the Pechukas-Yukawa formalism}

Depending on the nature of the noise, whether the source is longitudinal (with only diagonal elements) or transverse (with only off-diagonal elements), the system behaves differently. Longitudinal contributions result in decoherence in the system whereas transverse noise results in couplings to the environment\cite{PokSin, Wilson1}. Our analysis could be extended to various types of noise. For concreteness we consider a single composite source of longitudinal noise, $\delta h$ such that $\dot\delta h=\epsilon\eta M$. Here, $\eta$ is white noise, a random normal distributed stochastic process\cite{Kieselev, PokSin}, $M$ represents a general diagonal matrix and $\epsilon$ denotes the noise amplitude. For white noise, which is the formal derivative of a Wiener process, $W(t)$, the expectation is zero and the autocorrelation function is given by:

%\begin{widetext}
%\begin{eqnarray}
\begin{equation}
\begin{gathered}
\label{WhiteNoise}
\langle \eta_{mn}(\lambda), \eta_{mn}(\lambda^{'})\rangle= \delta(\lambda-\lambda^{'})\\
\langle \epsilon\eta_{mn}(\lambda), \epsilon\eta_{mn}(\lambda^{'})\rangle=\epsilon^2 \delta(\lambda-\lambda^{'})
\end{gathered}
\end{equation}
%\end{eqnarray}
%\end{widetext}

\noindent The correlation time $\tau_c = 0$.  
%This matrix is defined as $\dot\delta h=(\begin{matrix} a && 0 &&\dots \\ \vdots && \ddots &&\vdots\\ 0 && \dots &&  b \end{matrix})$ where $a$ and $b$ are constants with difference $\mu$. %For simplicity, we imposed that $\delta h \in \R^{NXN}; \delta h_{mn}=\delta h_{nm}$. This reduces the number of noise sources from $N^2$ to $\frac{N(N+1)}{2}$.  is characterised by its correlation time $\tau_c$, (the time between successive peaks in the noise) and its square-amplitude.  

Noise can break the degeneracy at level crossings, resulting in anticrossings. To ensure the applicability of the LZ model, we reduce the system from $N$ levels to 2. Again, under the assumption that levels outside the anticrossings are far away with weaker coupling interactions we show that the anticrossing is independent of all non-interacting level contributions. The Pechukas-Yukawa model is highly entangled, hence why it is important to verify that the conditions required for the LZ description are met. %The model described in Eq. (\ref{PNoise}), must be regarded as independent of all non-interacting levels about an anticrossing. 

Considering the stochastic Pechukas equations regarding the relative angular momenta, $l_{mn}$ described in Eq. (\ref{PNoise}), we obtain a driftless geometric Brownian motion for $\dot l_{mn}$ for levels $m$ and $n$ in an anticrossing. Then, in the region of the anticrossing $l_{mn}(\lambda)=l_{mn}(\lambda^{*}-\xi)e^{-\frac{\sigma^2}{2}(\lambda-(\lambda^{*}-\xi))+\sigma W(\lambda)}$, where $\sigma=\frac{\mu\epsilon}{\Delta_{min}}$. The start time of the levels approaching a minimum separation in a $\gamma$ neighbourhood of each other is taken as $(\lambda^{*}-\xi)$ as in Fig. 1, and $\mu$ denotes the difference in the noise components. The expectation of $l_{mn}$, $E(l_{mn})=l_{mn}(\lambda^{*}-\xi)$ and variance, $Var(l_{mn})=|l_{mn}(\lambda^{*}-\xi)|^2(e^{\frac{\sigma^2}{2}(\lambda-(\lambda^{*}-\xi))}-1)$. Here, $l_{mn}$ is a martignale, where in the long time limit, $l_{mn} \rightarrow 0$ with probability 1. Substituting these to determine the couplings between non-interacting levels we find that $l_{mi}$ and $l_{ni}$ are stochastic terms, where $Re(l_{mi})=0$ and $Re(l_{ni})=0$ with $Im(l_{mi}), Im(l_{ni})$ bound the interval $[-1, 1]$. Taking $\delta\lambda$ sufficiently small, these terms are negligible in the anticrossing. Applying this to the acceleration terms, we find that the difference in acceleration terms is also a stochastic term $\frac{4|l_{mn}|^2}{\Delta_{min}^3}$. Given that $\tau_{LZ}$ is short, the expectation is strongly bounded in a small interval, choosing $\delta\lambda$ sufficiently small, these terms are also negligible therefore linearising the level separations. Details are provided in Appendix C.  

In order for the LZ model to hold in the stochastic sense, it is necessary to consider anticrossings in a close vicinity of each other, such that they can be regarded as isolated crossings. The transition time of an anticrossing is changed under the influence of noise. Of particular interest are the influences of noise on the minimum separation. These in turn have an impact on both the probability of transitions and the transition times. 

%For Brownian noise, there is a uniform distribution of population with $P_{LZ}=\frac{1}{2}$. This describes a system that loses memory of its initial conditions\cite{Kieselev}.  In the stochastic sense, transition time changes depending on the type of noise being considered. 

 %For Brownian noise, $\tau_C \ll \tau_{LZ}$ and level repulsion tends to infinity, in agreement with the acceleration of levels given by $\dot v_m=\frac{2\beta^2}{\Delta_{min}^3}+2\frac{Re(\beta)\epsilon\eta_{mn}}{\Delta_{min}^2}$. Using this expression, we show the accelerations are solely dependent on transverse noise contributions. %the Langevin equation for Brownian noise is described as $\dot\delta h= -\tau_c\delta h+\epsilon\eta$,

\subsection{Isolated Crossings under the effects of noise}

Under the influences of noise on the minimum level separation at $\lambda^*$, we investigate its impact on the transition time to determine the conditions required to treat 2 nearby anticrossings independently. We consider 2 neighbouring anticrossings with minimum level separations at $\lambda^{*}$ and $\lambda^{**}=\lambda^*+\delta$ and transition times $\tau_{LZ}$ and $\tau^{'}_{LZ}$ respectively as in Fig. 1. The anticrossings are considered isolated given that the respective transition times do not overlap such that $(\lambda^{**}-\xi^{`})-(\lambda^*+\xi) > 0$. Then, the Landau-Zener transition model is applicable to describe the probabilities of population transitions.% In the adiabatic regime, $\tau_{LZ}=\frac{\Delta_{min}}{\dot\lambda}$. We take symmetric level (avoided) crossings such that $\tau_{LZ}=2\xi$. 

We denote the distance between levels $d(\lambda) = x_m-x_n$, where $x_m>x_n$, $m$ and $n$ label the levels involved at an anticrossing. Expanding $d(\lambda)$ about $\lambda^*$, where $f(\lambda)=-\frac{\sigma^2}{2}(\lambda-(\lambda^{*}-\xi))+\sigma\eta(\lambda-(\lambda^{*}-\xi))$, we obtain the following (details are provided in Appendix E):

%\begin{widetext}
%\begin{eqnarray}
\begin{equation}
\begin{gathered}
\label{ExpandGap}
d(\lambda) = \Delta_{min}\\
+\delta\lambda^2\left(\frac{4|l_{mn}(\lambda^{*}-\xi)|^2}{\Delta_{min}^3}e^{2f(\lambda^{*})}+\epsilon\mu\dot\eta(\lambda^{*})\right)
%d(\lambda) = \Delta_{min}+\delta\lambda^2(\frac{4\beta^2}{\Delta_{min}^3}+\epsilon\dot\eta_{mm}-\epsilon^{'}\dot\eta_{nn})-\sigma^2\xi+2\sigma\eta(\xi)}
%d(\lambda) = \Delta_{min}+\delta\lambda^2(\frac{4\beta^2}{\Delta_{min}^3}+\tau_c\Omega+\epsilon\dot\eta_{mm}-\epsilon^{'}\dot\eta_{nn})
\end{gathered}
\end{equation}
%\end{eqnarray}
%\end{widetext}

\noindent Let $d(\lambda^*+\xi)=\gamma$, then one obtains an expression for the minimum separation:

%\begin{widetext}
%\begin{eqnarray}
\begin{equation}
\begin{gathered}
\label{GapNoise}
\Delta_{min}=\gamma-\\
\xi^2\left(\frac{4|l_{mn}(\lambda^{*}-\xi)|^2}{\Delta_{min}^3}e^{2f(\lambda^{*})}+\epsilon\mu\dot\eta(\lambda^{*})\right)
%\Delta_{min}=\gamma-\xi^2(\frac{4\beta^2}{\Delta_{min}^3}+\epsilon\dot\eta_{mm}-\epsilon^{'}\dot\eta_{nn})
%\Delta_{min}=\gamma-\xi^2(\frac{4\beta^2}{\Delta_{min}^3}+\tau_c\Omega+\epsilon\dot\eta_{mm}-\epsilon^{'}\dot\eta_{nn})
\end{gathered}
\end{equation}
%\end{eqnarray}
%\end{widetext}

\noindent This describes the relationship between the minimum separation and the difference in noise terms, where $\Delta_{min}\geq 0$. These effects  on the level separation affect $\tau_{LZ}$ in the same way. When $\mu=0$, $d(\lambda) = \Delta_{min}+\delta\lambda^2(\frac{4|l_{mn}(\lambda^{*}-\xi)|^2}{\Delta_{min}^3}e^{2f(\lambda^{*})})$. Then the conditions for an isolated anticrossing resemble that of the deterministic case.  % We observe that only longitudinal noise affects the minimum gap, . Eq.(\ref{GapNoise}) describes the most general case for 2 different sources of Brownian noise with different characteristics such that $\eta_{mm}\neq\eta_{nn}$.  

%For multiple sources of white noise with the same characteristics such that $\dot\delta h_{mn}=\epsilon\eta_{mn}$, then $d(\lambda)$ expanded about $\lambda^{*}$, $d(\lambda)=\Delta_{min}+\delta\lambda^2(\ddot\delta h_{mm}-\ddot\delta h_{nn}+\frac{4\beta^2}{\Delta_{min}^3})$. Hence the minimum level separation is given by the following:%Eq.(\ref{GapNoise}) becomes:

%\begin{widetext}
%\begin{eqnarray}
%\begin{equation}
%\begin{gathered}
%\label{GapNoise}
%\Delta_{min}=\gamma-\xi^2(\frac{4\beta^2}{\Delta_{min}^3}+\epsilon(\dot\eta_{mm}-\dot\eta_{nn}))
%\Delta_{min}=\gamma-\xi^2(\frac{4\beta^2}{\Delta_{min}^3}+\tau_c\Omega+\epsilon\dot\eta_{mm}-\epsilon^{'}\dot\eta_{nn})
%\end{gathered}
%\end{equation}
%\end{eqnarray}
%\end{widetext}

Using Eq.(\ref{GapNoise}) in the bound for the transition times, one obtains the following bound, dependent of the difference between the noise sources at a single anticrossing (all details provided in Appendix E):

%\begin{widetext}
%\begin{eqnarray}
\begin{equation}
\begin{gathered}
\label{NoiseBoundStrict}
\eta(\lambda^{*})>\frac{1}{\xi\epsilon\mu}(\gamma-2\dot\lambda(\delta-\xi^{`})-\\
\frac{\xi}{\epsilon\mu}\left(\frac{4|l_{mn}(\lambda^{*}-\xi)|^2}{\Delta_{min}^3}e^{2f(\lambda^{*})}\right).%\frac{4}{\Delta_{min}^3}|l_{mn}(\lambda^{*}-\xi)|^2e^{\frac{\sigma^2}{2}\xi})
\end{gathered}
\end{equation}
%\end{eqnarray}
%\end{widetext}

\noindent Given this bound is satisfied, the 2 anticrossings are independent of each other, as such the LZ model is applicable. %As $\eta$ is Gaussian distributed, one can express this bounds probabilistically. %, let $A=\int{\dot\eta_{mm}-\dot\eta_{nn}}d\lambda$  describing a Gaussian distribution and $\mu=\frac{1}{\epsilon}\int{\frac{2\dot\lambda(\delta-\xi^{`})-\gamma}{\xi^2}-\frac{4\beta^2}{\Delta_{min}^3}}d\lambda$, then one obtains the following:

%\begin{widetext}
%\begin{eqnarray}
%\begin{equation}
%\begin{gathered}
%\label{WhtNoiseBound}
%P(A< \mu )=\frac{1}{\mu\sqrt{2\pi}}\int^{A(\lambda+\mu)}_{A(\lambda-\mu)} e^-\frac{(A(\lambda)-A(\lambda^*))^2}{2\mu}dA
%\end{gathered}
%\end{equation}
%\end{eqnarray}
%\end{widetext}

%In the most general case with multiple sources of noise with different characteristics, we obtain the following bound:

%\begin{widetext}
%\begin{eqnarray}
%\begin{equation}
%\begin{gathered}
%\label{NoiseBoundStrict}
%\int{\epsilon\dot\eta_{mm}-\epsilon^{'}\dot\eta_{nn}}d\lambda < \int{\frac{2\dot\lambda(\delta-\xi^{`})-\gamma}{\xi^2}+\frac{4\beta^2}{\Delta_{min}^3}}d\lambda
%\end{gathered}
%\end{equation}
%\end{eqnarray}
%\end{widetext}

%This follows the same probability of occurence given by the gaussian distribution in Eq.(\ref{WhtNoiseBound}) with $A=\int{\epsilon\dot\eta_{mm}-\epsilon^{'}\dot\eta_{nn}}d\lambda$, and $\mu=\int{\frac{2\dot\lambda(\delta-\xi^{`})-\gamma}{\xi^2}+\frac{4\beta^2}{\Delta_{min}^3}}d\lambda$ (all details provided in Appendix B). This determines the probability of this bound being satisfied under a Gaussian distribution. Hence, 
Therefore, we have shown from the analysis of the levels at a level crossing or anticrossing, the conditions the LZ model imposes on the Pechukas-Yukawa, under the influence of noise. % modelrequired for the Landau-Zener model hold under the Pechukas formalism. Additionally, we have determined the conditions .   % can apply probabilistic arguments using the Gaussian distribution, giving:

\section{Discussion and Conclusions.}
\noindent  We investigated the compatibility of the LZ transition model in the Pechukas-Yukawa formalism. Taking as starting point all the assumptions that form the basis of the LZ model, we explored the conditions they impose on the Pechukas-Yukawa formalism to be applicable. This led to the development of the understanding of level crossings and anticrossings under this setting, identifying various properties of the level interaction. Particularly, we provided a detailed insight on the level repulsions extended to the influence of external noise and its impacts on the minimum separations characterising anticrossings. 

The investigation of level repulsions at an anticrossing under the influence of longitudinal noise was not possible  without a thorough description of the level dynamics given by the Pechukas-Yukawa formalism. From this, we  built on prior works by [\onlinecite{Zagoskin}] and [\onlinecite{Wilson1}], to gain insight on the level interactions beyond the LZ probability. Under this description, one could investigate the differences in scaling properties observed between edge and intermediate state transitions, observed in [\onlinecite{Zagoskin}] and [\onlinecite{Wilson1}]. An attractive development to this investigation would be to apply the LZ transition probability to the Pechukas-Yukawa description of quantum states which could lead to the exploration of quantum phase transitions through the initial conditions of the eigenvalues of a quantum Hamiltonian system. The eigenstate coefficients have been expressed using the Pechukas equations such that one could extend this description to obtain both the occupation dynamics and the coherences of the system, crucial to the development of AQ-this leads us into our future works. An interesting extension to these works would be to consider the effects of different types of noise such as coloured noise and the impacts of transverse components.
%We also considered the effects of a single source of composite longitudinal external noise in the system. This description allowed for the investigation of level repulsions at an anticrossing, not possible We%, providing more detail on the level (avoided) crossing, the conditions required to apply the Landau-Zener model and the types of noise it is capable of describing such that level interactions are considered well separated. 

Additionally, these results can be used as a a starting point, to gain insight on multistate LZ transitions. The standard LZ model deals only with the 2 interacting levels. Extending to the multistate problem could yield more interesting physics analytics. The Pechukas-Yukawa model concerns an interacting system of $N$ entangled levels. It is highly equipped to consider interacting systems with entangled states. In further works it would be useful to consider the detailed analytics of multiple level interactions and their influence on each other's dynamics. One could extend this description to determine the impacts of noise using a master equation. 

%under a density matrix description would allow one to further evaluate both the occupation dynamics and the coherences of a system. This holds interests in the dynamics of adiabatic systems and sources of decoherence, crucial to the development of adiabatic quantum computers. %where level (avoided) crossings occur, we investigated the conditions required to apply the Landau-Zener 2-level  model such that we can to make use of the Landau-Zener proabability of remaining in a state after a level interaction. This probability approximated the occupation dynamics of the system. % Under a density matrix setting, one may consider the occupation dynamics and evolution of coherences independent of the conditions imposed on the system by the Landau-Zener model. This allows for an analytical description of an N level interacting system. 

\section*{Acknowledgments} We are grateful to Sergey Savel'ev, Alexander Veselov, Anatoly Nieshtadt, Huaizhong Zhou and Chunrong Feng for the valuable discussions that greatly improved the manuscript. This work has been supported by EPSRC through the grant No. EP/M006581/1.

\section*{Appendix A: Reducing System Levels Down to 2-Level Crossings}

It is shown below that when there is a level crossing, all non-interacting levels are considered far apart. Then, the Pechukas equations can be reduced to only the interacting levels. %This enables the Pechukas-Yukawa formalism to applied in the Landau-Zener model. 

Suppose $x_m=x_n$ are the interacting levels and all other levels are far apart i.e $x_m-x_k$ and $x_n-x_k$ large for $k \neq n,m$ and angular moment $l_{mk} l_{kn}$ are small. Then, the quotient is small and so one takes the folllowing approximation

%\begin{widetext}
%\begin{eqnarray}
\begin{equation}
\begin{gathered}
\label{lmnCross}
\dot l_{mn}=\sum_{k\neq m,n}{l_{mk}l_{kn}\left(\frac{1}{{(x_m-x_k)}^2}-\frac{1}{({x_k-x_n)}^2}\right)} \approx 0 \\
\end{gathered}
\end{equation}
%\end{eqnarray}
%\end{widetext}
By the definition of the Pechukas equations when $x_m=x_n$, $l_{mn}=0$ hence $l_{mn}$ stays constantly zero throughout the transition time.

Similarly, the other non-interacting angular momenta can be paired into the following coupled differential equations.  All other terms are negligible. These are approximated as follows: for $i \neq m, n$

%\begin{widetext}
%\begin{eqnarray}
\begin{equation}
\begin{gathered}
\label{lmnCross2}
\dot l_{mi}%=\sum_{k\neq m,i}{l_{mk}l_{ki}\left(\frac{1}{{(x_m-x_k)}^2}-\frac{1}{({x_k-x_i)}^2}\right)} \\
%\approx {l_{mn}l_{ni}\left(\frac{1}{{(x_m-x_n)}^2}-\frac{1}{({x_n-x_i)}^2}\right)} \\
\approx  {l_{mn}l_{ni}\left(\frac{1}{{(x_m-x_n)}^2}\right)}\\
\dot l_{ni}\approx  {l_{nm}l_{mi}\left(\frac{1}{{(x_n-x_m)}^2}\right)}\\
\end{gathered}
\end{equation}
%\end{eqnarray}
%\end{widetext}

%Consider an $N$ level system, with levels $x_m$ and $x_n$ crossing such that $x_m=x_n$ and $l_{mn}=0$ where all other level separations in contrast are negligible with $x_m-x_i=0$ where $i \in \mathbb{N}, i \neq n$. In order to justify the application of the pechukas-Yukawa formalism to the Landau-Zener model, it is neccessary that the Pechukas equations can b reduced to soleley the interacting levels. For the relative angular momenta terms, of particular interest are $l_{mn}, l_{mi}$ and $l_{nj}$ where  $j \in \mathbb{N}, j \neq m$ where we have the following:

%\begin{widetext}
%\begin{eqnarray}
%\begin{equation}
%\begin{gathered}
%\label{lmnCross}
%\dot l_{mn}=\sum_{k\neq m,n}{l_{mk}l_{kn}\left(\frac{1}{{(x_m-x_k)}^2}-\frac{1}{({x_k-x_n)}^2}\right)}\\
%\dot l_{mi}=\sum_{k\neq m,i; i\neq n}{l_{mk}l_{ki}\left(\frac{1}{{(x_m-x_k)}^2}-\frac{1}{({x_k-x_n)}^2}\right)}\\+{l_{mn}l_{ni}\left(\frac{1}{{(x_m-x_n)}^2}-\frac{1}{({x_n-x_i)}^2}\right)}
%\end{gathered}
%\end{equation}
%\end{eqnarray}
%\end{widetext}

%\noindent $l_{nj}$ takes the same form as $l_{mi}$ and the same arguments follow. For $l_{mn}$ the levels are far away and so $\left(\frac{1}{{(x_m-x_k)}^2}-\frac{1}{({x_k-x_n)}^2}\right)$ is negligible hence $\dot l_{mn} \approx 0$ where $l_{mn}=0$ and constant. For $l_{mi}$ the terms under the sum are negligible, with the only significant term being ${l_{mn}l_{ni}\left(\frac{1}{{(x_m-x_n)}^2}\right)}$. 

Applying l'Hopital's rule on this term twice, we have shown this term tends to 0 as $\lambda \rightarrow \lambda^{*}$ demonstrating the relative angular momenta terms can be reduced to only the interacting levels under this approximation. It follows that the acceleration terms are also independent of all other level interactions, determined by the following:

%\begin{widetext}
%\begin{eqnarray}
\begin{equation}
\begin{gathered}
\label{vmCross}
\dot v_m=2\sum_{i\neq n}{\frac{{{|l}_{mi}|}^2}{{(x_m-x_i)}^3}}+{\frac{{{|l}_{mn}|}^2}{{(x_m-x_n)}^3}}\\
\dot v_i=2\sum_{i, j\neq m, n}{\frac{{{|l}_{ij}|}^2}{{(x_i-x_j)}^3}}+{\frac{{{|l}_{mj}|}^2}{{(x_m-x_j)}^3}}\\
+{\frac{{{|l}_{nj}|}^2}{{(x_n-x_j)}^3}}\\
\end{gathered}
\end{equation}
%\end{eqnarray}
%\end{widetext}

\noindent Again the same argument holds for $\dot v_n$ as does $\dot v_m$. Using the expressions in Eq.(\ref{lmnCross}), $l_{mi}$ is constant hence the terms under the sum in $\dot v_m$ are negligible. After performing l'Hopital 3 times, the expression ${\frac{{{|l}_{mn}|}^2}{{(x_m-x_n)}^3}}$ was found to tend to 0 as $\lambda \rightarrow \lambda^{*}$. Expanding about $\lambda^{*}$, level separation is described by $x_m-x_n=\delta\lambda(v_m-v_n)+\delta\lambda^2(\dot v_m-\dot v_n)+\O(\delta\lambda^3)$, where acceleration terms idependently tend to 0 at a level crossing. This linearises level separtions in this region during the Landau-Zener transition. For $v_m=v_n$, the numerator and denominator in the accelleration terms, identically go to 0, thus one can treat $\dot v_m$ as constant, such that for small $\delta\lambda$ level separation can be taken as linear. This argument holds identically for $\dot v_n$.  For the $\dot v_i$ expression, all terms are negligible. %demonstrating $\dot v_m$ can indeed be treated as independent of all other level interactions. 

This demonstrates the applicability of the Pechukas-Yukawa formalism to the LZ model as one can indeed reduce and $N$ level system down to 2, neglecting all other interactions.

\section*{Appendix B: Reducing $N$ Levels Down to 2-Anticrossings Without Noise}

Anticrossings occur when levels approaching each other, reach a local minimum before deflecting away. In such cases, $x_m-x_n=\Delta_{min}$ and $l_{mn}$ is not neccessarily 0. In the same way, Eq.(\ref{lmnCross}) and Eq.(\ref{vmCross}) apply. Under the same approximation that all other levels are far away, again $\dot l_{mn}=0$ thus $l_{mn}=\beta$ where $\beta$ is a constant. Considering the equations for $l_{mi}$ and $l_{ni}$, the only surviving terms are:

%\begin{widetext}
%\begin{eqnarray}
\begin{equation}
\begin{gathered}
\label{lmnAvoid}
\dot l_{mi}={l_{mn}l_{ni}\left(\frac{1}{{(x_m-x_n)}^2}\right)}={l_{ni}\beta\left(\frac{1}{{\Delta_{min}}^2}\right)}\\
\dot l_{ni}={l_{mn}l_{mi}\left(\frac{1}{{(x_n-x_m)}^2}\right)}=-{l_{mi}\beta^{*}\left(\frac{1}{{\Delta_{min}}^2}\right)}
\end{gathered}
\end{equation}
%\end{eqnarray}
%\end{widetext}

We obtain coupled differential equations. Rewritten as $\left(\begin{matrix} \dot l_{mi}\\ \dot l_{ni} \end{matrix}\right)=\frac{1}{\Delta_{min}^2}\left(\begin{matrix} 0 && \beta \\ -\beta^{*} && 0 \end{matrix}\right)\left(\begin{matrix} l_{mi}\\  l_{ni} \end{matrix}\right)$. The system is readily solved as:% Diagonalising the matrix and changing bases to the eigenvectors, we can simply integrate the decoupled set of equations. We obtain the following:

%\begin{widetext}
%\begin{eqnarray}
\begin{equation}
\begin{gathered}
\label{lmnCouple}
l_{mi}=\frac{i\beta}{|\beta|}\frac{1}{2}\left(e^{\frac{i|\beta|}{\Delta_{min}^2}}+e^{\frac{-i|\beta|}{\Delta_{min}^2}}\right)
=\frac{i\beta}{|\beta|}\cos\left(\frac{|\beta|}{\Delta_{min}^2}\right)\\
l_{ni}=\frac{-1}{2}(e^{\frac{i|\beta|}{\Delta_{min}^2}}-e^{-\frac{i|\beta|}{\Delta_{min}^2}})
=-i\sin(\frac{|\beta|}{\Delta_{min}^2})
%l_{mi}\dot l_{mi}=\frac{-\beta}{\beta^{*}}l_{ni}\dot l_{ni}
\end{gathered}
\end{equation}
%\end{eqnarray}
%\end{widetext}

Then about $\lambda^{*}$, the relative angular momenta $l_{mn}$ are constants independent of all other levels. We further showed, $l_{mi}$ and $l_{ni}$ are constants with $Re(l_{mi})=0$ and $Re(l_{ni})=0$ with $Im(l_{mi})$ and $Im(l_{ni})$, bounded between [-1, 1], hence the couplings between the levels involved in an anticrossing and those that are not, are weak . This allows for treating the anticrossing, independent of all other levels. Substituting these results into Eq.(\ref{vmCross}), $\dot v_i=0$, the only surviving terms in $\dot v_m$ and $\dot v_n$ are constants; $(\dot v_m-\dot v_n)=\frac{{4|\beta|}^2}{{\Delta_{min}}^3}$. For sufficiently small $\delta\lambda$, one can linearise the level separations such that level evolutions are reduced to only the interacting levels. Then, it is justifiable in applying the Pechukas-Yukawa formalism to the Landau-Zener model for anticrossings. %Furthermore, we note that under the Landau-Zener model, all other non-interacting level relative angular momenta take the same value for all points in $\tau_{LZ}$.

%Rearranging for $\frac{1}{{\Delta_{min}}^2}$ and substituting into the other equations, we obtain: 

%\begin{widetext}
%\begin{eqnarray}
%\begin{equation}
%\begin{gathered}
%\label{lmnCouple}
%l_{mi}\dot l_{mi}=\frac{-\beta}{\beta^{*}}l_{ni}\dot l_{ni}
%\end{gathered}
%\end{equation}
%\end{eqnarray}
%\end{widetext}

%Integrating these expressions with respect to $\lambda$, we obtain the following relationship between the coupled equations:

%\begin{widetext}
%\begin{eqnarray}
%\begin{equation}
%\begin{gathered}
%\label{lmnSolve}
%l_{mi}=i\sqrt{\frac{-\beta}{\beta^{*}}}l_{ni}
%\end{gathered}
%\end{equation}
%\end{eqnarray}
%\end{widetext}

%Substituting this expression back into Eq.(\ref{lmnAvoid}), we obtain the following:

%\begin{widetext}
%\begin{eqnarray}
%\begin{equation}
%\begin{gathered}
%\label{lmnSolve}
%l_{mi}=e^{-i\frac{|\beta|\lambda}{{\Delta_{min}}^2}}=l_{ni}
%\end{gathered}
%\end{equation}
%\end{eqnarray}
%\end{widetext}  rather constants with $|l_{mi}|=1$. 

\section*{Appendix C: Isolated Crossings for a Deterministic Case}

\noindent We denote level separations as $d(\lambda)=x_1-x_2$, where $d(\lambda^{*})=\Delta_{min}$. Let $\delta\lambda=\lambda-\lambda^{*}$, then expanding about $\lambda^{*}$, $d(\lambda)=\Delta_{min}+\delta\lambda(v_1-v_2+\dot\delta h_{11}-\dot\delta h_{22})+\delta\lambda^2(\frac{4\beta^2}{\Delta_{min}^3}+\ddot\delta h_{11}-\ddot\delta h_{22})+\O(\delta\lambda^3)$. Given that $d(\lambda)$ reaches a local minimum at $\lambda^{*}$, then $v_1-v_2+\dot\delta h_{11}-\dot\delta h_{22}=0$. 

\noindent Take $d(\lambda^{*}+\xi)=\gamma$, such that one could rearrange the equation to obtain:

%\begin{widetext}
%\begin{eqnarray}
\begin{equation}
\begin{gathered}
\label{Min}
\Delta_{min}=\gamma-\xi^2\frac{4\beta^2}{\Delta_{min}^3}
%\Delta_{min}=\gamma-\xi^2(\frac{4\beta^2}{\Delta_{min}^3}+\tau_c\Omega+\epsilon\dot\eta_{11}-\epsilon^{'}\dot\eta_{22})
\end{gathered}
\end{equation}
%\end{eqnarray}
%\end{widetext}

\noindent In order to ensure that anticrossings can be treated independently, $(\lambda^{**}-\xi^{'})-(\lambda^{*}+\xi)>0$ where $\lambda^{**}=\lambda^{*}+\delta$. Recall $\tau_{LZ}=\frac{\Delta_{min}}{\dot\lambda}=2\xi$ for a symmetric anticrossing. Then it is essentially $\tau_{LZ}<2(\delta-\xi^{'})$. One could rearrange this bound for $\Delta_{min}$, 

%\begin{widetext}
%\begin{eqnarray}
\begin{equation}
\begin{gathered}
\label{NoiseBoundStrict}
\gamma-\xi^2\frac{4\beta^2}{\Delta_{min}^3}<2\dot\lambda(\delta-\xi^{'})
\end{gathered}
\end{equation}
%\end{eqnarray}
%\end{widetext}

%\begin{widetext}
%\begin{eqnarray}
%\begin{equation}
%\begin{gathered}
%\label{NoiseBoundStrict}
%\int{\epsilon\dot\eta_{mm}-\epsilon^{'}\dot\eta_{nn}}d\lambda < \int{\frac{2\dot\lambda(\delta-\xi^{`})-\gamma+\xi^2\tau_c\Omega+\frac{4\beta^2}{\Delta_{min}^3}}{\xi^2}}d\lambda
%\end{gathered}
%\end{equation}
%\end{eqnarray}
%\end{widetext}

Given that $\delta>\frac{1}{2\dot\lambda}(\gamma-\frac{4\beta^2}{\Delta_{min}^3}\xi^2)+\xi^{`}$, the conditions for anticrossings to be treated independently are satisfied. 

\section*{Appendix D: Reducing $N$ Levels Down to 2-Anticrossings With Noise}

When noise is present in a system, level interactions are always non-degenerate occuring with anticrossings. To determine the applicability of the Pechukas-Yukawa formalism under dissiptive influences, it is neccessary to ensure that level interactions in an anticrossing are independent of all other interactions. Again, $x_m-x_n=\Delta_{min}$ at some $\lambda^*$ (denoting the point of minimum separation) and $l_{mn}$ is not neccessarily 0. Similarly to Eq.(\ref{lmnCross}), we have the following for the coupling between levels at an anticrossing. 

%\begin{widetext}
%\begin{eqnarray}
\begin{equation}
\begin{gathered}
\label{lmnAvoid}
\dot l_{mn}=\sum_{k\neq m,n}{l_{mk}l_{kn}\left(\frac{1}{{(x_m-x_k)}^2}-\frac{1}{({x_k-x_n)}^2}\right)}\\
+\frac{(x_m-x_n)(l_{mk}\dot\delta h_{km}-\dot\delta h_{mk}l_{kn})}{(x_m-x_k)(x_n-x_k)}+\\
\dot\delta h_{mn}(v_m-v_n)+\frac{l_{mn}(\dot\delta h_{mm}-\dot\delta h_{nn})}{(x_m-x_n)}\\
\end{gathered}
\end{equation}
%\end{eqnarray}
%\end{widetext}

We consider a single source of composite longitudinal noise. Again, we assume all non-interacting levels are far away with weak couplings such that $l_{mk} l_{kn}$ are small for $k \neq n,m$. This simplifies the relative angular momena dynamics to the following:%where all other levels are far away again:

%\begin{widetext}
%\begin{eqnarray}
\begin{equation}
\begin{gathered}
\label{lmnAvoid}
\dot l_{mn}\approx \frac{l_{mn}(\dot\delta h_{mm}-\dot\delta h_{nn})}{(x_m-x_n)}\\
\approx\frac{l_{mn}}{\Delta_{min}}\epsilon\mu\eta
\end{gathered}
\end{equation}
%\end{eqnarray}
%\end{widetext}

\noindent where $\epsilon$ denotes the noise amplitude, $\mu$ is a constant giving the difference between the noise components and $\eta$  represents a white noise stochasti term. Let $\sigma=\frac{\epsilon\mu}{\Delta_{min}}$. We consider separately, real and imaginary components. In each component, we observe a driftless geometric Brownian motion. 

%\begin{widetext}
%\begin{eqnarray}
\begin{equation}
\begin{gathered}
\label{lmnAvoid}
\dot Re(l_{mn})=\sigma Re(l_{mn})\eta\\
\dot Im(l_{mn})=\sigma Im(l_{mn})\eta\\
\end{gathered}
\end{equation}
%\end{eqnarray}
%\end{widetext}

Using the Euler-Maruyama method to solve these stochastic differential equations, we rewrite the expression for $Re(l_{mn})$ as $dRe(l_{mn})=\sigma Re(l_{mn})dW$. Integrating these terms, where we zero out noise at $\lambda^{*}-\xi$, we obtain the following:

%\begin{widetext}
%\begin{eqnarray}
\begin{equation}
\begin{gathered}
\label{lmnAvoid}
\int_{\lambda^{*}-\xi}^{\lambda}\frac{dRe(l_{mn})}{l_{mn}}=\sigma dW\\
\end{gathered}
\end{equation}
%\end{eqnarray}
%\end{widetext}

Applying Ito's formula, 

%\begin{widetext}
%\begin{eqnarray}
\begin{equation}
\begin{gathered}
\label{Ito}
d(Ln(Re(l_{mn})))=\frac{dRe(l_{mn})}{l_{mn}}-\frac{1}{2}\frac{1}{Re(l_{mn})^2}dRe(l_{mn})dRe(l_{mn})\\
\end{gathered}
\end{equation}
%\end{eqnarray}
%\end{widetext}

\noindent where $dRe(l_{mn})dRe(l_{mn})$ is the quadratic variation of the stochastic differential equation such that $dRe(l_{mn})dRe(l_{mn})=\sigma^2Re(l_{mn})^2d\lambda$. Substituting this into the integral, we have:

%\begin{widetext}
%\begin{eqnarray}
\begin{equation}
\begin{gathered}
\label{lmnAvoid}
\int_{\lambda^{*}-\xi}^{\lambda}d(Ln(Re(l_{mn})))+\frac{\sigma^2}{2}=\sigma dW\\
\end{gathered}
\end{equation}
%\end{eqnarray}
%\end{widetext}

Then, 

%\begin{widetext}
%\begin{eqnarray}
\begin{equation}
\begin{gathered}
\label{lmnAvoid}
Ln\left(\frac{Re(l_{mn}(\lambda))}{Re(l_{mn}(\lambda^{*}-\xi))}\right)\\
=-\frac{1}{2}\sigma^2(\lambda-(\lambda^{*}-\xi))+\sigma W(\lambda)
\end{gathered}
\end{equation}
%\end{eqnarray}
%\end{widetext}

Exponentiating the result, we find that $Re(l_{mn}(\lambda))=Re(l_{mn}(\lambda^{*}-\xi))e^{-\frac{\sigma^2}{2}(\lambda-(\lambda^{*}-\xi))+\sigma W(\lambda)}$. Using the same method to solve for the imaginary components, we have $Im(l_{mn}(\lambda))=Im(l_{mn}(\lambda^{*}-\xi))e^{-\frac{\sigma^2}{2}(\lambda-(\lambda^{*}-\xi))+\sigma W(\lambda)}$. Combining these terms, $l_{mn}(\lambda)=l_{mn}(\lambda^{*}-\xi)e^{-\frac{\sigma^2}{2}(\lambda-(\lambda^{*}-\xi))+\sigma\eta(\lambda)}$ in the region of the transition time. This term has expectation, $E(l_{mn})=l_{mn}(\lambda^{*}-\xi)$ and variance $Var(l_{mn})=|l_{mn}(\lambda^{*}-\xi)|^2(e^{\frac{\sigma^2}{2}(\lambda-(\lambda^{*}-\xi))}-1)$. Here, $(\lambda^{*}-\xi)$ represents the start time of levels approaching a minimum separation in a $\gamma$ neighbourhood of each other. This describes $l_{mn}$ as a martingale where for $\lambda\rightarrow \infty, l_{mn} \rightarrow 0$ with probability 1, which follows from the law of iterative logarithm. %For $\mu\epsilon \in (0,1)$, noise dominates despite being weak however has a small variance, then on averge it remains constant. 

The equations for  $l_{mi}$ are given by the following:

%\begin{widetext}
%\begin{eqnarray}
\begin{equation}
\begin{gathered}
\label{lmnAvoid}
\dot l_{mi}=\sum_{k\neq m,i; i \neq n}{l_{mk}l_{ki}\left(\frac{1}{{(x_m-x_k)}^2}-\frac{1}{({x_k-x_i)}^2}\right)}\\
+\frac{(x_m-x_i)(l_{mk}\dot\delta h_{km}-\dot\delta h_{mk}l_{ki})}{(x_m-x_k)(x_i-x_k)}+\\
\dot\delta h_{mn}(v_m-v_i)+\frac{l_{mi}(\delta h_{mm}-\delta h_{ii})}{(x_m-x_i)}+\\
{l_{mk}l_{kn}\left(\frac{1}{{(x_m-x_k)}^2}-\frac{1}{({x_k-x_n)}^2}\right)}\\
+\frac{(x_m-x_n)(l_{mk}\dot\delta h_{km}-\dot\delta h_{mk}l_{kn})}{(x_m-x_k)(x_n-x_k)}+\\
\dot\delta h_{mn}(v_m-v_n)+\frac{l_{mn}(\delta h_{mm}-\delta h_{nn})}{(x_m-x_n)}\\
\end{gathered}
\end{equation}
%\end{eqnarray}
%\end{widetext}

The equations are identical for $l_{ni}$. Again, under the same assumptions used for $\dot l_{mn}$, we obtain the following pairs of coupled differential equations:% and coupling the  we have: Assume levels not involved in the avoided crossing are far away and their relative angular momenta in comparison weak. Hence, the only non-negligible relative angular momenta terms between levels not involved in the avoided crossing are given by the following. One can pair these terms to obtain coupled differential equations.%

%\begin{widetext}
%\begin{eqnarray}
\begin{equation}
\begin{gathered}
\label{lmnAvoid}
\dot l_{mi}\approx l_{mn}l_{ni}\left(\frac{1}{{(x_m-x_n)}^2}\right)= l_{mn}l_{ni}\left(\frac{1}{{\Delta_{min}}^2}\right)\\
\dot l_{ni}\approx l_{nm}l_{mi}\left(\frac{1}{{(x_n-x_m)}^2}\right)= -l_{mn}^{*}l_{mi}\left(\frac{1}{{\Delta_{min}}^2}\right)\\
\end{gathered}
\end{equation}
%\end{eqnarray}
%\end{widetext}

%Again we obtain coupled equations between $l_{mi}$ and $l_{ni}$. 

Taking a matrix of ordinary differential equations, we have 

%\begin{widetext}
%\begin{eqnarray}
\begin{equation}
\begin{gathered}
\label{CoupleAvoid}
\left(\begin{matrix} \dot l_{mi}\\ \dot l_{ni} \end{matrix}\right)=\frac{f(\lambda)}{\Delta_{min}^2}\left(\begin{matrix} 0 && l_{mn}(\lambda^{*}-\xi) \\ -l_{mn}^{*}(\lambda^{*}-\xi) && 0 \end{matrix}\right)\left(\begin{matrix} l_{mi}\\  l_{ni} \end{matrix}\right)
\end{gathered}
\end{equation}
%\end{eqnarray}
%\end{widetext}

Where $f(\lambda)=e^{\frac{\sigma^2}{2}(\lambda-(\lambda^{*}-\xi))+\sigma\eta(\lambda)}$, capturing the stochastic element. Diagonalising the matrix and changing bases to the eigenvectors, we can simply integrate the decoupled set of equations. We obtain the following:

%\begin{widetext}
%\begin{eqnarray}
\begin{equation}
\begin{gathered}
\label{lmnCouple}
l_{mi}=\frac{il_{mn}}{2|l_{mn}|}\left(e^{i\frac{f(\lambda)}{\Delta_{min}^2}|l_{mn}|}+e^{-i\frac{f(\lambda)}{\Delta_{min}^2}|l_{mn}|}\right)\\
=\frac{il_{mn}}{|l_{mn}|}cos\left(\frac{f(\lambda)}{\Delta_{min}^2}|l_{mn}|\right)\\
l_{ni}=-\frac{1}{2}\left(e^{i\frac{f(\lambda)}{\Delta_{min}^2}|l_{mn}|}-e^{-i\frac{f(\lambda)}{\Delta_{min}^2}|l_{mn}|}\right)\\
=-isin\left(\frac{f(\lambda)}{\Delta_{min}^2}|l_{mn}|\right)\\
%l_{mi}\dot l_{mi}=\frac{-\beta}{\beta^{*}}l_{ni}\dot l_{ni}
\end{gathered}
\end{equation}
%\end{eqnarray}
%\end{widetext}

Then, $l_{mi}$ and $l_{ni}$ are stochastic terms, where $Re(l_{mi})=0$ and $Re(l_{ni})=0$ with $Im(l_{mi}), Im(l_{ni})$ bounded in the interval $[-1, 1]$. Taking $\delta\lambda$ sufficiently small, these terms are negligible in the anticrossing. Applying these relative angular momenta formulae to the acceleration terms (again modelling the noise to be a longitudinal composite source) we have the following: %These stochastic equations are bounded in [-i, i], hence their respective motions are small.

%\begin{widetext}
%\begin{eqnarray}
\begin{equation}
\begin{gathered}
\label{vmAvoid}
\dot v_m=2\sum_{i\neq n}{\frac{{{|l}_{mi}|}^2}{{(x_m-x_i)}^3}}+\frac{2\dot\delta h_{mi}Re(l_{mi})}{{(x_m-x_i)}^2}\\
+{\frac{{{|l}_{mn}|}^2}{{(x_m-x_n)}^3}}+\frac{2\dot\delta h_{mn}Re(l_{mn})}{{(x_m-x_n)}^2}\\
\dot v_i=2\sum_{i, j\neq m, n}{\frac{{{|l}_{ij}|}^2}{{(x_i-x_j)}^3}}+\frac{2\dot\delta h_{mi}Re(l_{mi})}{{(x_m-x_i)}^2}\\
+{\frac{{{|l}_{mj}|}^2}{{(x_m-x_j)}^3}}+\frac{2\dot\delta h_{mi}Re(l_{mi})}{{(x_m-x_i)}^2}\\
+{\frac{{{|l}_{nj}|}^2}{{(x_n-x_j)}^3}}+\frac{2\dot\delta h_{mi}Re(l_{mi})}{{(x_m-x_i)}^2}\\
\end{gathered}
\end{equation}
%\end{eqnarray}
%\end{widetext}

\noindent All terms are negligible for $v_i$ under the approximation on the level separation in this regions is negligible. 

For the difference between $\dot v_m$ and $\dot v_n$, all terms under the sum are negligible except for ${\frac{4{|l_{mn}|}^2}{{\Delta_{min}}^3}}$, which is independent of all other levels. To determine the effects of the stochastic terms on the difference between accelerations, we consider the expectation during $\tau_{LZ}$. The expectation of $|l_{mn}|^2$, is given by:

%\begin{widetext}
%\begin{eqnarray}
\begin{equation}
\begin{gathered}
\label{vmAvoid}
|l_{mn}|^2=Re(l_{mn})^2+Im(l_{mn})^2\\
E|l_{mn}|^2=E(Re(l_{mn})^2)+E(Im(l_{mn})^2)\\
=Var(Re(l_{mn}))+Var(Im(l_{mn}))\\
+E^2(Re(l_{mn}))+E^2(Im(l_{mn}))\\
%=(Re(l_{mn}(\lambda^{*}+\xi))^2+Im(l_{mn}(\lambda^{*}+\xi))^2)(1+e^{\frac{\sigma^2}{2}(\lambda-(\lambda^{*}-\xi))}-1)\\
=|l_{mn}(\lambda^{*}+\xi)|^2e^{\frac{\sigma^2}{2}(\lambda-(\lambda^{*}-\xi))}
\end{gathered}
\end{equation}
%\end{eqnarray}
%\end{widetext}

Then the expectation of the difference between the acceleration terms are given by $\frac{4|l_{mn}(\lambda^{*}+\xi)|^2}{\Delta_{min}^3}e^{\frac{\sigma^2}{2}(\lambda-(\lambda^{*}-\xi))}$. These dynamics are bounded between $[\frac{4|l_{mn}(\lambda^{*}+\xi)|^2}{\Delta_{min}^3}, \frac{4|l_{mn}(\lambda^{*}+\xi)|^2}{\Delta_{min}^3}e^{\xi\sigma^2}]$ where $\lambda \in [\lambda^{*}-\xi, \lambda^{*}+\xi]$. For $\tau_{LZ}$ being short time durations, this motion is under stricter bounds, near-constant. In taking $\delta\lambda$ small enough, the difference in accelaration terms are negligible, linearising the level separations. Then it is observed that indeed the Pechukas-Yukawa formalism under the influence of noise is applicable to the LZ model, reducing the system from N levels to 2.

\section*{Appendix E: Isolated Crossings for a Stochastic Case}
 
Under the influences of noise, the level separations expanded about $\lambda^{*}$ are given by $d(\lambda) = \Delta_{min}+\delta\lambda^2\left(\frac{4|l_{mn}(\lambda^{*}-\xi)|^2}{\Delta_{min}^3}e^{2f(\lambda^{*})}+\epsilon\mu\dot\eta(\lambda^{*})\right)$. Let us denote the level separation at the final instant by $d(\lambda^{*}+\xi)=\gamma$, then one can once again rearrange for $\Delta_{min}$: %$d(\lambda)=\Delta{min}+\delta\lambda^2(\frac{4|l_{mn}|^2}{\Delta_{min}^3}+\epsilon\mu\dot\eta(\lambda^{*})$

%\begin{widetext}
%\begin{eqnarray}
\begin{equation}
\begin{gathered}
\label{Min}
\Delta_{min}=\gamma-\\
\xi^2\left(\frac{4|l_{mn}(\lambda^{*}-\xi)|^2}{\Delta_{min}^3}e^{2f(\lambda^{*})}+\epsilon\mu\dot\eta(\lambda^{*})\right)
%\Delta_{min}=\gamma-\xi^2(\frac{4|l_{mn}((\lambda^{*})|^2}{\Delta_{min}^3}+\epsilon\mu\dot\eta(\lambda^{*}))
%\Delta_{min}=\gamma-\xi^2(\frac{4\beta^2}{\Delta_{min}^3}+\tau_c\Omega+\epsilon\dot\eta_{11}-\epsilon^{'}\dot\eta_{22})
\end{gathered}
\end{equation}
%\end{eqnarray}
%\end{widetext}
 
 Again, for symmetric anticrossings, one obtains the following bound:
 
%\begin{widetext}
%\begin{eqnarray}
\begin{equation}
\begin{gathered}
\label{Min}
\gamma-\xi^2\left(\frac{4|l_{mn}(\lambda^{*}-\xi)|^2}{\Delta_{min}^3}e^{2f(\lambda^{*})}+\epsilon\mu\dot\eta(\lambda^{*})\right)<2\dot\lambda(\delta-\xi^{`})
%\Delta_{min}=\gamma-\xi^2(\frac{4\beta^2}{\Delta_{min}^3}+\tau_c\Omega+\epsilon\dot\eta_{11}-\epsilon^{'}\dot\eta_{22})
\end{gathered}
\end{equation}
%\end{eqnarray}
%\end{widetext}

Integrating over the transition time on both sides and rearranging for $\eta(\lambda^{*})$, we reduce the bound to the following:

%\begin{widetext}
%\begin{eqnarray}
\begin{equation}
\begin{gathered}
\label{Min}
\eta(\lambda^{*})>\frac{1}{\xi\epsilon\mu}(\gamma-2\dot\lambda(\delta-\xi^{`})-\\
\frac{\xi}{\epsilon\mu}\left(\frac{4|l_{mn}(\lambda^{*}-\xi)|^2}{\Delta_{min}^3}e^{2f(\lambda^{*})}\right)\\
%\eta(\lambda^{*})>\frac{\xi}{\epsilon\mu}(\frac{1}{\xi^2}(\gamma-2\dot\lambda(\delta-\xi^{`})-\frac{4|l_{mn}((\lambda^{*})|^2}{\Delta_{min}^3})%+\epsilon\mu\dot\eta(\lambda^{*}))<2\dot\lambda(\delta-\xi^{`})
%\Delta_{min}=\gamma-\xi^2(\frac{4\beta^2}{\Delta_{min}^3}+\tau_c\Omega+\epsilon\dot\eta_{11}-\epsilon^{'}\dot\eta_{22})
\end{gathered}
\end{equation}
%\end{eqnarray}
%\end{widetext}
 
This provides a bound on the system, accounting for noise. Given the noise at $\lambda^{*}$ satisfies this bound, the conditions for level crossings to be treated independently hold. Then, the LZ model is applicable. These equations detail a system with a single composite source of longitudinal noise and its impact on the probability of isolated level crossings. These can be explored for various cases under different types of noise. %simpler cases when the characteristics are the same or when there is simply a single source of noise on the system. 


\begin{thebibliography}{32}

\bibitem{Zagoskin1}
A.M. Zagoskin, E. Il'ichev, M. Grajcar, J.J. Betouras, and F. Nori, Front. Physics {\bf 2}, 33 (2014).

\bibitem{Requist4}
R Requist, J Schliemann, AG Abanov, D Loss, Phy. Rev. B {\bf 71}, 115315 (2005).

\bibitem{Fahri1}
E. Fahri et al., Science {\bf 292}, 472 (2001).

\bibitem{Fahri2}
A. M. Childs, E. Farhi, J. Preskill, Phys. Rev. A {\bf 65}, 012322 (2001).

\bibitem{Sarovar}
M. Sarovar, K. C. Young, New Journal of Physics, {\bf 15} 125032 (2013).

\bibitem{Candia}
R. Di Candia, B. Mejia, H. Castillo, J. S. Pedernales, J. Casanova and E. Solano, Phys. Rev. Lett. {\bf 111}, 240502 (2013).

%\bibitem{Feynman}
%R. P. Feynman, International Journal of Theoretical Physics, {\bf 21} 6/7, (1982)

\bibitem{Pechukas}
P. Pechukas, Phys. Rev. Lett. {\bf 51}, 943 (1983)

\bibitem{Yukawa3}
T. Yukawa and T. Ishikawa, Prog. Theor. Phys. Suppl. {\bf 98}, 157 (1989).

\bibitem{Gernot}
G. Schaller, S. Mostame, R. Schutzhold, Phys. Rev. A {\bf 73}, 062307 (2006).

\bibitem{Haake}
F. Haake, {\it Quantum Signitures of Chaos}, Ch. 6 (Springer, Berlin, 2001).

\bibitem{Ours}
M. A. Qureshi, J. Zhong, J. J. Betouras, A. M. Zagoskin, Phys. Rev. A {\bf 95}, 032126 (2017).

\bibitem{Requist}
R. Requist, Phys. Rev. A {\bf 86} (2), 022117 (2012).

\bibitem{Zagoskin}
A. M. Zagoskin, S. Savel'ev and F. Nori, Phys. Rev. Lett. {\bf 98}, 120503 (2007).

%\bibitem{Love}
%J. D. Biamonte and P. J. Love, Phys. Rev. A {\bf 78}, 012352 (2008).

%\bibitem{Caneva}
%T. Caneva, R. Fazio and G. E. Santoro, J. Phys.: Conf. Ser. {\bf 143}, 012004 (2009).

\bibitem{Berends}
R. Barends, A. Shabani, L. Lamata, J. Kelly {\it et al}, Nature {\bf 534},  17658 (2016).

\bibitem{Huyghebaert}
J. Huyghebaert and H. De Raedt, J. Phys. A {\bf 23}, 5777-5793 (1990).

%\bibitem{Poulin}
%D. Poulin, A. Qarry, R. D. Somma and F. Verstraete, Phys. Rev. Lett. {\bf 106}, 170501 (2011).

\bibitem{Wilson1}
R. D. Wilson, A. M. Zagoskin and S. Savel'ev, Phys. Rev. A {\bf 82}, 052328 (2010).

\bibitem{Ours2}
M. A. Qureshi, J. Zhong, Z. Qureshi, P. Mason, J. J. Betouras, A. M. Zagoskin, Phys. Rev. A {\bf 97}, 032117 (2018).

\bibitem{Kayanuma}
Y. Kayanuma, H. Nakayama, Phys. Rev. B {\bf 57}, 13099 (1998)

\bibitem{PokSin}
V. L. Pokrovsky, N. A. Sinitsyn, Phys. Rev. B {\bf 67}, 144303 (2003)

\bibitem{Cejnar}
P. Cejnar, J. Jolie, Prog. Part. Nucl. Phys. {\bf 62}, 210-256 (2009)

\bibitem{Leggett}
A. J. Leggett, S. Chakravarty, A. T. Dorsey, M. P. A. Fisher, A. Garg and W. Zwerger, Rev. Mod. Phys. {\bf 59}, 1 (1987)

\bibitem{Yukawa1}
T. Yukawa, Phys. Rev. Lett. {\bf 54}, 1883 (1985).

\bibitem{Wilson2}
R. D. Wilson, A. M. Zagoskin, S. Savel'ev,  M. J. Everitt and F. Nori, Phys. Rev. A {\bf 86},  052306 (2012).

\bibitem{Vitanov}
N. V. Vitanov and B. M. Garraway, Phys. Rev. A {\bf 53}, 4288 (1996)

\bibitem{Kleppner}
J. R. Rubbmark, M. M. Kash, M. G. Littman and D. Kleppner, Phys. Rev. A {\bf 23}, (6) 3107 (1981)

\bibitem{Wilkinson1}
M. Wilkinson, J. Phys. A: Math. Gen. {\bf 21}, 4021-4037 (1988)

\bibitem{Wilkinson2}
M. Wilkinson, J. Phys. A: Math. Gen. {\bf 22}, 2795-2805 (1989)

\bibitem{Zakrezewski}
J. Zakrzewski, D. Delande, M. Kus, Phys. Rev. E {\bf 47}, (3) 1665 (1993)

\bibitem{Amin2}
M. H. S. Amin, D. V. Averin and J. A. Nesteroff, Phys. Rev. A {\bf 79}, 022107 (2009)

\bibitem{Kieselev}
M. B. Kenmoe, H. N. Phien, M. N. Kiselev and L. C. Fai, Phys. Rev. B {\bf 87}, 224301 (2013)

\bibitem{Luo}
Z. X. Luo and M. E. Raikh, Phys. Rev. B {\bf 95}, 064305 (2017)

\bibitem{Gangardt}
G. E. Astrakharchik, D. M. Gangardt, Yu. E. Lozovik, I. A. Sorokin, Phys. Rev. E {\bf 74}, 021105 (2006)

%\bibitem{Kato}
%T. Kato, Journal of the Physical Society of Japan. {\bf 5} (6): 435–439 (1950)

%\bibitem{Yukawa2}
%T. Yukawa, Phys. Lett. {\bf 116A}, 227 (1986).

%\bibitem{Blanes}
%S. Blanes, F. Casas, J. Oteo, and J. Ros. 
%Magnus and Fer expansions for matrix differential equations: the convergence problem. 
%J. Phys. A, {\bf 31}, 259 (1998).

%\bibitem{Gantmacher}
%F. R. Gantmacher. The theory of matrices. Chelsea Publishing Co., New
%York, 1959. Two volumes. Translated by K. A. Hirsch.

%\bibitem{Oteo}
%S. Klarsfeld and J. A. Oteo. 
%The Baker-��Campbell-��Hausdorff formula and the convergence of the Magnus expansion. 
%J. Phys. A, {\bf 22}, 4565 (1989).

%\bibitem{Magnus}
%W. Magnus. 
%On the exponential solution of differential equations for a linear operator. 
%Comm. Pure and Appl. Math., {\bf 7}, 639 (1954).

%\bibitem{Light}
%P. Pechukas and J. C. Light. 
%On the exponential form of the time displacement operator in quantum mechanics. 
%J. Chem. Phys., {\bf 7}, 3897 (1966).

%\bibitem{Chowski}
%S. Wojciechowski, Phys. Lett. {\bf 111A}, 3 (1985).

%\bibitem{YukawaLax}
%T. Yukawa, Phys. Lett. A, {\bf 116}, 5 (1986).


%\bibitem{Yvon}
%J. Yvon, {\it La Theorie Statistque des Fluids et L'equation d'Etat}, Act. Scient. et ind. No. 203 (Hermann, Paris, 1935) .

%\bibitem{Balescu}
%R. Balescu, {\it Equilibrium and Nonequillibrium Statistical Mechanics}, Vols 1/2 (Wiley, New York, 1975). 

%\bibitem{Fern}
%F. M. Fern´andez. Convergence of the Magnus expansion. Phys. Rev. A,
%41(5):2311–2314, 1990.

%\bibitem{Stockmann}
%H. J. Stockmann, {\it Eigenvalue Dynamics in Quantum Chaos An Introduction}, (Cambridge Press, 1999).

%\bibitem{Alexeev}
%B. V. Alexeev, {\it Generalized Boltzmann Physical Kinetics} (Elsevier, 2004).

%\bibitem{Requist1}
%R. Requist, Phys. Rev. A {\bf 86} (2), 022117 (2012).

%\bibitem{Suzuki}
%S. Suzuki, J. Phys.: Conf. Ser. {\bf 302}, 012046 (2011).

%\bibitem{Requist2}
%R. Requist, O. Pankratov, Phys. Rev. A {\bf 83} (5), 052510 (2011).

%\bibitem{Requist3}
%R Requist, O Pankratov, Physical Review A {\bf 81}, 042519 (2010).

%\bibitem{Sergey}
%S. Savel'ev, F. Marchesoni and F. Nori, Phys. Rev. E {\bf 70}, 061107 (2004).

%\bibitem{PBS}
%M. Baake and U. Schlegel, Proc. Steklov Inst. Math. {\bf 275}, 167(2011).

\end{thebibliography}
\end{document}